%% file: main.tex
\documentclass[conference, 10pt]{IEEEtran}

\input{new_commands}

\begin{document}

\title{Understanding the Effects of Real-World Behavior in Statistical Disclosure Attacks}

\author{\IEEEauthorblockN{Simon Oya\IEEEauthorrefmark{1}$^1$,
Carmela Troncoso\IEEEauthorrefmark{2}$^2$ and
Fernando P{\'e}rez-Gonz{\'a}lez\IEEEauthorrefmark{1}\IEEEauthorrefmark{2}$^3$}
\IEEEauthorblockA{\IEEEauthorrefmark{1}Signal Theory and Communications Dept., University of Vigo\\
Vigo 36310, Spain\\
$^1$\texttt{simonoya@gts.uvigo.es}\qquad $^3$\texttt{fperez@gts.uvigo.es}} %\{simonoya,fperez\}@gts.uvigo.es}}
\IEEEauthorblockA{\IEEEauthorrefmark{2}Gradiant (Galician R\&D Center in Advanced Telecommunications)\\
Vigo 36310, Spain\\
$^2$\texttt{ctroncoso@gradiant.org}}}

\maketitle
\input{sections/abstract}
\input{sections/intro}
\input{sections/LSDA}

\input{sections/realworld}

\input{sections/analysis}

\input{sections/conclusions}

{\small \section*{ACKNOWLEDGEMENT}

This work was partially funded by the Spanish Government and the ERDF under project TACTICA, by the Spanish Government under project COMPASS (TEC2013-47020-C2-1-R), by the Galician Regional Government and the ERDF under projects Consolidation of Research Units (GRC2013/009), REdTEIC (R2014/037) and AtlantTIC, and by the EU 7th Framework Programme (FP7/2007-2013) under grant agreements 610613 (PRIPARE) and 285901 (LIFTGATE).}

\bibliographystyle{IEEEtran}
\bibliography{IEEEabrv,references}

\end{document}

%% file: new_commands.tex
\usepackage{amssymb}
\usepackage{graphicx}

\usepackage{url}
\usepackage[normalem]{ulem}
\usepackage{booktabs} % To add toprule and other stuff to tables.
\usepackage{amsmath,amsfonts,amssymb}
\usepackage{subfig}
\usepackage{multirow}

\usepackage[usenames,dvipsnames]{color}

\usepackage{xspace}
\newcommand{\Enron}[0]{{\em Email}\xspace}
\newcommand{\Gowalla}[0]{{\em Location}\xspace}
\newcommand{\MailingList}[0]{{\em MailingList}\xspace}

\usepackage{xifthen}% provides \isempty test

\newcommand{\Rx}[0]{\mathbf{R}_x}
\newcommand{\Rxyx}[0]{\mathbf{R}_{xyx}}

\newcommand{\pj}[1]{\mathbf{p}_{#1}}
\renewcommand{\P}[0]{\mathbf{P}}
\newcommand{\Pest}[0]{\hat{\mathbf{P}}}

\newcommand{\pjest}[1]{\hat{\mathbf{p}}_{#1}}
\newcommand{\pest}[2]{\hat{p}_{#1,#2}}
\newcommand{\U}[0]{\mathbf{U}}
\newcommand{\Y}[0]{\mathbf{Y}}
\newcommand{\yj}[1]{\mathbf{y}_{#1}}
\newcommand{\Yj}[1]{\mathbf{Y}_{#1}}
\newcommand{\Exp}[1]{\text{E}\left\{ #1 \right\}}
\newcommand{\Var}[1]{\text{Var}\left\{ #1 \right\}}
\newcommand{\Cov}[1]{\text{Cov}\left\{ #1 \right\}}

\newcommand{\Sj}[0]{\mathbf{S}_j}

\newcommand{\prob}[2]{\ensuremath{p_{#1,#2}}}
\newcommand{\sendprof}[1]{\ensuremath{\mathbf{q}_{#1}}}

\newcommand{\uniformi}[1]{\ensuremath{\upsilon_{#1}}}

\newcommand{\MSEi}[0]{\ensuremath{\mbox{MSE}_i}}

\newcommand{\MSEimax}[0]{\ensuremath{\mbox{MSE}_i^{+}}}
\newcommand{\MSEimin}[0]{\ensuremath{\mbox{MSE}_i^{-}}}

\newcommand{\M}[0]{\ensuremath{\mathbf{M}}}
\newcommand{\MM}[0]{\ensuremath{\mathbf{M}_2}}
\newcommand{\MMM}[0]{\ensuremath{\mathbf{M}_3}}
\newcommand{\MMMM}[0]{\ensuremath{\mathbf{M}_4}}

\newcommand{\muv}[0]{\ensuremath{\boldsymbol\mu}}

%% file: sections/abstract.tex
\begin{abstract}
%\boldmath
High-latency anonymous communication systems prevent passive eavesdroppers from inferring communicating partners with certainty. However, disclosure attacks allow an adversary to recover users' behavioral profiles when communications are persistent. Understanding how the system parameters affect the privacy of the users against such attacks is crucial. Earlier work in the area analyzes the performance of disclosure attacks in controlled scenarios, where a certain model about the users' behavior is assumed. In this paper, we analyze the profiling accuracy of one of the most efficient disclosure attack, the least squares disclosure attack, in realistic scenarios. We generate real traffic observations from datasets of different nature and find that the models considered in previous work do not fit this realistic behavior. We relax previous hypotheses on the behavior of the users and extend previous performance analyses, validating our results with real data and providing new insights into the parameters that affect the protection of the users in the real world.
\end{abstract}

% keywords
\begin{IEEEkeywords} anonymity, mixes, performance analysis \end{IEEEkeywords}

% \fi
\IEEEpeerreviewmaketitle

%% file: sections/intro.tex
\section{Introduction}
Mixes aim at providing anonymity in communication networks by acting as routers that hide the correspondence between senders and receivers of messages. 
These anonymous communication channels operate by gathering the messages they receive, changing their appearance cryptographically and outputting them in batches, in what are called \emph{rounds} of mixing. However, providing perfect anonymity through mixes is not possible in practice, due to constraints in the bandwidth of the communication channel and the delay tolerated by users. Because of this, an adversary observing the system in the long-term may infer the frequency with which a certain sender communicates with a certain receiver by means of a \emph{disclosure attack}~\cite{SDA,MD04,PMDA,Vida,PETS12}. One of these strategies, called the Least Squares Disclosure Attack (LSDA) \cite{PETS12,WIFS12,PETS14}, has been proven to outperform previous statistical variants~\cite{GlobalSIP} while keeping its computational cost much lower than more sophisticated approaches, such as~\cite{Vida}. One advantage of LSDA is that it is particularly suitable for analysis, due to the availability of closed-form expressions for its prediction error in terms of the system parameters. Such performance analysis is of paramount importance since it helps the designer of mix-based anonymous communication systems to understand how to improve the protection of the users.

Previous works analyze the prediction error of LSDA in mix-based systems~\cite{PETS12,WIFS12,PETS14,GlobalSIP} under specific assumptions on the users' and mix behavior. However, these results have only been confirmed by computer-generated observations and therefore it is not clear whether they apply in real-world scenarios. In this document, we delve into how users behave in reality. We gather data from real databases of different nature, which we then use to show that previous analyses of the attack fall short when tested against real data. We analyze the hypotheses that are needed for the performance analysis of LSDA to be applicable in real-world scenarios and develop a new generalized closed-form expression for the attacker's error when estimating the relationships between users in mixes, which we then evaluate with real traffic.
Real-world datasets have been used in other works to compare between different disclosure attacks~\cite{WPES13} or to analyze the properties of real traffic~\cite{malinka2009analyses}. Our approach is different, as we are interested in understanding the effects of real-world user behavior on the performance of the least squares disclosure attack.

The document is structured as follows: we describe the least squares attack in the following section, together with the system model and notation we use in the paper. In Sect.~\ref{sec:realworld}, we study the statistical properties of real-world behavior in our system. We carry out and evaluate a new performance analysis of LSDA in Sect.~\ref{sec:analysis}, and conclude in Sect.~\ref{sec:conclusions}.

%% file: sections/LSDA.tex
\section{The Least Squares Disclosure Attack}
\label{sec:LSDA}

\begin{figure}[t]
\begin{center}
\noindent
  \includegraphics[width=2.3in]{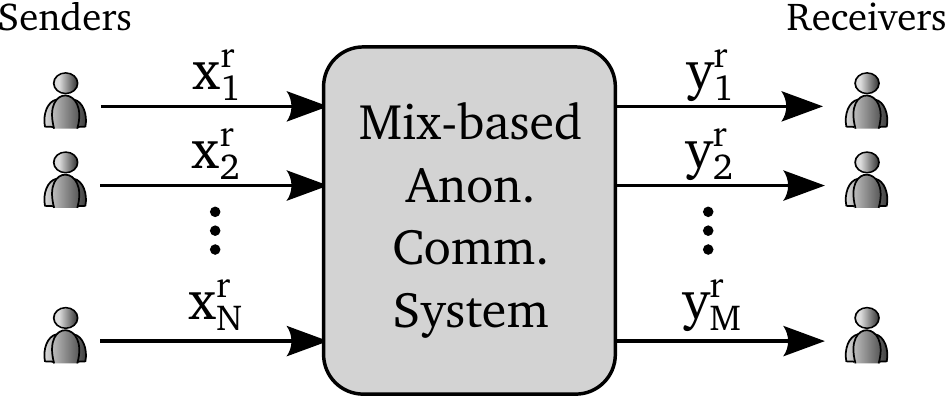}
  \caption{System model during the communication \emph{round} $r$.}\label{fig:sysmodel}
\end{center}
\end{figure}

The Least Squares Disclosure Attack (LSDA), introduced by P{\'e}rez-Gonz{\'a}lez and Troncoso in \cite{PETS12}, estimates the intensity of the communication between each sender-receiver pair in a mix-based anonymous channel by solving a least squares problem. This intensity is represented by the \emph{transition probabilities} $\prob{j}{i}$, which model the \emph{average probability} that a message sent by sender $i\in\{1,2,\cdots,N\}$ is addressed to receiver $j\in\{1,2,\cdots,M\}$. These probabilities are commonly grouped per sender in the so-called \emph{sending profiles}, $\sendprof{i}\doteq[\prob{1}{i},\cdots,\prob{M}{i}]^T$. An attacker that observes the number of messages sent and received during $\rho$ communication rounds obtains the LSDA estimator by solving
\begin{equation} \label{eq:LSDA}
\Pest=\left(\U^T\U\right)^{-1} \U^T \Y\,,
\end{equation}
where $\Pest$ is a $N\times M$ matrix containing the estimation of the transition probability $\pest{j}{i}$ in its $i,j$th entry, $\U$ is a $\rho\times N$ matrix containing the amount of messages sent by sender $i$ in round $r$, denoted $x_i^r$, in its $r,i$th entry, and $\Y$ is a $\rho\times M$ matrix with the number of messages received by receiver $j$ in round $r$, denoted $y_j^r$, in its $r,j$th entry. Figure~\ref{fig:sysmodel} shows an example of the system and notation employed.
The estimator in \eqref{eq:LSDA} was proven to be unbiased and asymptotically efficient, in the sense that its variance approaches zero as the length of the observation window $\rho$ increases~\cite{PETS12,PETS14}, in mix-based systems where all messages leave the mix in each round.

Denoting the $j$th column of $\P$ by $\pj{j}\doteq[\prob{j}{1},\cdots,\prob{j}{N}]^T$, and the $j$th column of $\Y$ by $\yj{j}\doteq[y_j^1,\cdots,y_j^\rho]$, \eqref{eq:LSDA} can be decoupled as
\begin{equation} \label{eq:LSDAj}
 \pjest{j}=\left(\U^T\U\right)^{-1} \U^T \yj{j}\,.
\end{equation}
This latter formulation is specially useful to carry out a performance analysis of the attack.

%% file: sections/realworld.tex
\section{Modeling Real-World Behavior}
\label{sec:realworld}

In this section, we study real-world user behavior from observations generated with real traffic, showing that previous performance analyses of LSDA are not valid in this scenario because the assumptions they are based on are rather unrealistic. We propose alternative hypotheses that are adequate to model real-world user behavior, which we then use in Sect.~\ref{sec:analysis} to assess the performance of the LSDA estimator.

\subsection{Generating real-world observations}
\label{sec:realworld.datasets}

In order to analyze real-world behavior, we have chosen to generate observations by taking real traffic from datasets of different nature, whose users could have relied on mix-based systems to enhance their privacy, and anonymize this traffic using different mix configurations.
We work with three datasets, whose basic information is summarized in Table~\ref{tab:info_datasets}:
\begin{enumerate}
 \item {\bf \Enron:} This dataset contains around $220\,000$ emails sent from $294$ different email addresses, which have been extracted from the Enron corpus.\footnote{\url{http://www.cs.cmu.edu/~./enron/}} Messages with multiple recipients are treated as different messages sent simultaneously, one for each recipient.
 \item {\bf \Gowalla:} This dataset contains around $400\,000$ location check-ins taken from the $500$ most active users of Gowalla social networking website.\footnote{\url{http://snap.stanford.edu/data/loc-gowalla.html}} Users checking-in are considered as the senders, while the locations form the set of receivers. We consider only the $500$ most active users for computational reasons: LSDA works with large-size matrices which grow with the number of senders and receivers of the system.
 \item {\bf \MailingList:} we have processed the public mailing lists of Indimedia,\footnote{\url{http://lists.indymedia.org/}} obtaining almost $180\,000$ messages from the $500$ most active senders. Each mailing list is considered as a receiver, while users posting to these mailing lists are senders.
\end{enumerate}

 \begin{table}[t!]
 \begin{center}
   \caption{Basic information of the datasets.}
  \label{tab:info_datasets}
  \begin{tabular}{ r | r r r r }
    Dataset 	& No.~messages 	& Duration (hours) & Senders 	& Receivers 	\\ \hline
    \Enron 	& $220\,032$ 	& $32\,416.8$ 	& 294 		& $17\,017$ 	\\
%    \Gowalla 	& $2\,815\,449$	& $4\,344.0$ 	& $64\,225$  	& $1\,013$ 	\\
    \Gowalla & $406\,484$ 	&$4\,344.0$	& 500		& 559 \\
%    \MailingList & $293\,414$  	& $76\,239.8$	& $28\,237$  	& 693   \\
    \MailingList & $178\,937$ & $76166.4$	& 500		& 510
  \end{tabular}
  \end{center}
\end{table}

We anonymize the traces from these datasets using two types of mixes, which differ in the event that triggers the flushing of messages:
\begin{enumerate}
 \item {\bf Threshold mix:} this mix gathers messages until it has stored $t$ of them, and then forwards each one to its correspondent recipient.
 \item {\bf Timed mix:} this mix stores the messages it receives and, after a period of time $\tau$, outputs each one to their recipients.
\end{enumerate}

To generate the adversary's observations, we choose values of $t$ and $\tau$ that provide an acceptable degree of anonymity while keeping the delay of messages under a reasonable bound. We adopt the following criteria: in the threshold mix, we choose a value $t=100$ and, in the timed mix, we select values of $\tau$ to ensure that $\approx 100$ messages are mixed on average per round, while also considering that a delay of more that $24$ hours is intolerable for users. This makes $\tau=12$ hours for \Enron, $\tau=1$ hour for \Gowalla and $\tau=24$ hours for \MailingList, with an average of $\approx 100$ messages per round in the first two, and $\approx 57$ in the latter. The result of this anonymization is a set of observations from $\{X_i^r\}$ and $\{Y_j^r\}$.

\subsection{Modeling the input process}
\label{sec:realworld.input}

The input process, $\{X_i^r\}$, which models the amount of messages from each user arriving to the mix in each round, is determined by the frequency with which users send messages and by the firing condition of the mix. When the anonymization channel is a threshold mix, previous analyses~\cite{PETS12,WIFS12,GlobalSIP} assume that the input process follows a \emph{multinomial distribution}, and, when the channel is a timed mix, authors in \cite{PETS14} assume that the number of messages  each user sends to the mix can be independently modeled as a \emph{Poisson process}. 

\begin{figure*}[!t]
  \centering
  \subfloat[\Enron, threshold mix, $t=100$.]{\includegraphics[width=2.3in]{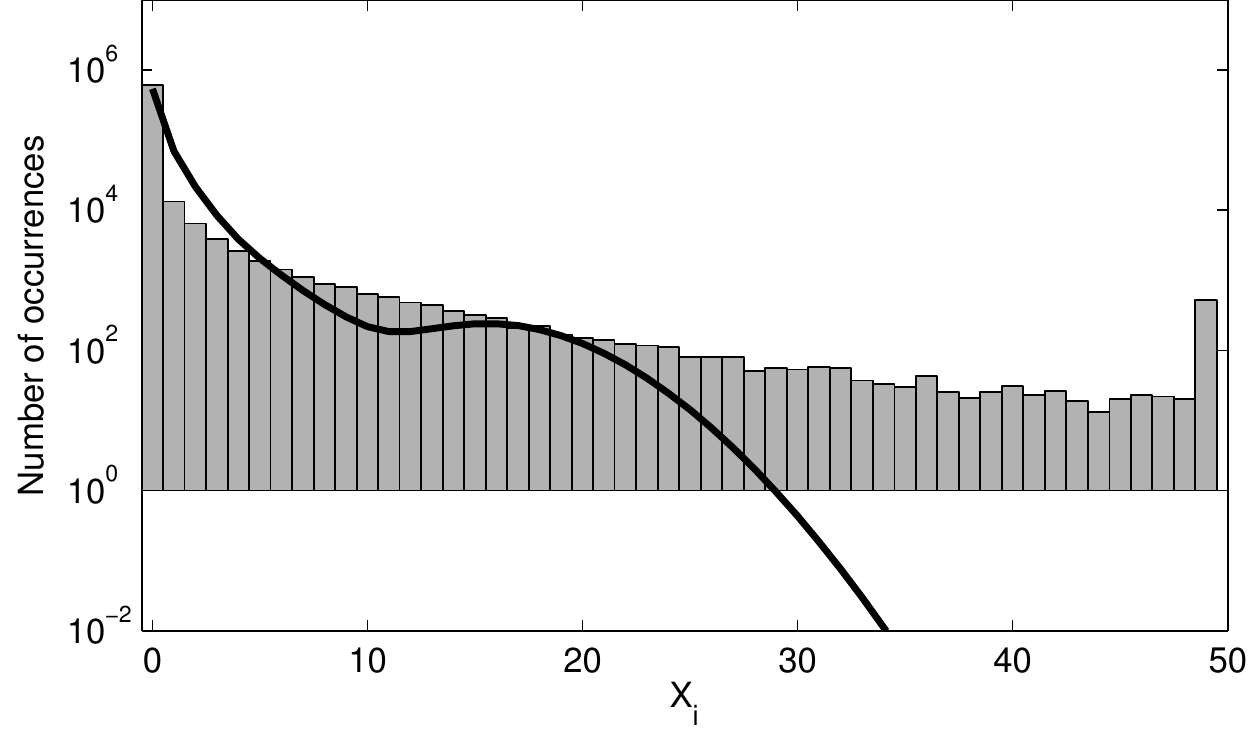} \label{fig:Hist_EN_t100}}
  \hfil
  \subfloat[\Gowalla, threshold mix, $t=100$.]{\includegraphics[width=2.3in]{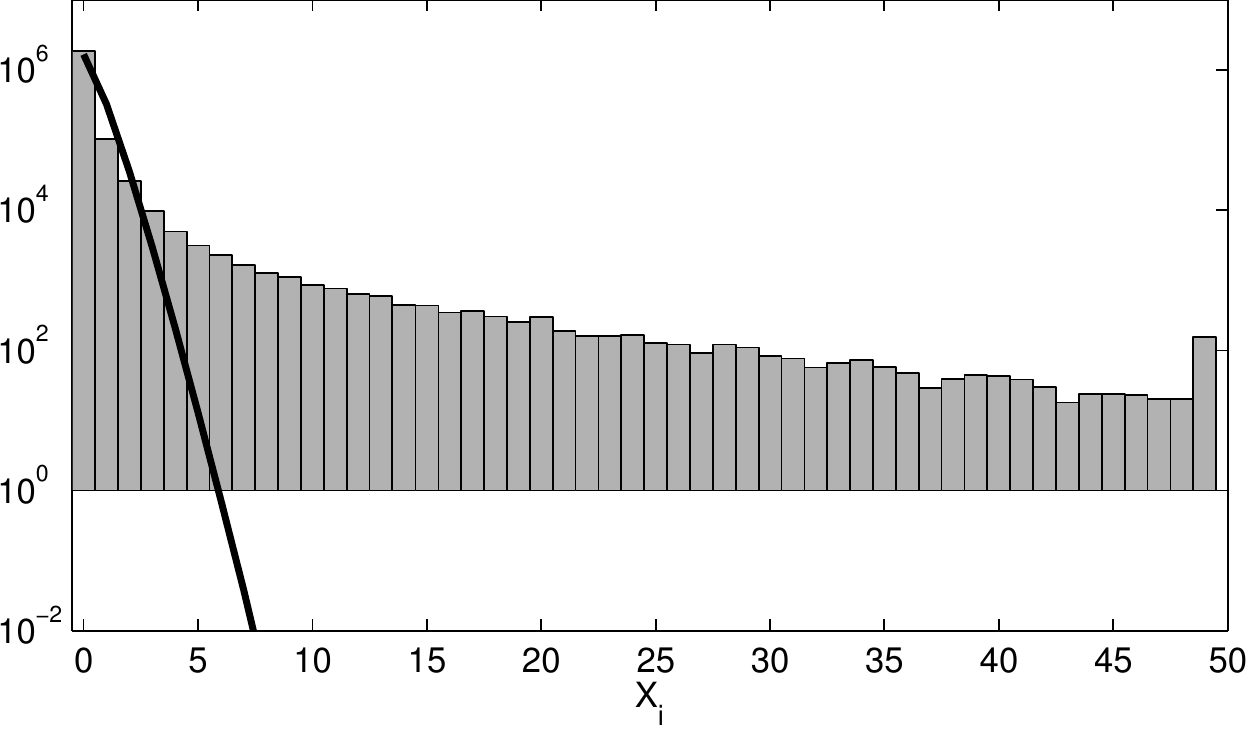} \label{fig:Hist_GO_t100}}
  \hfil
  \subfloat[\MailingList, threshold mix, $t=100$.]{\includegraphics[width=2.3in]{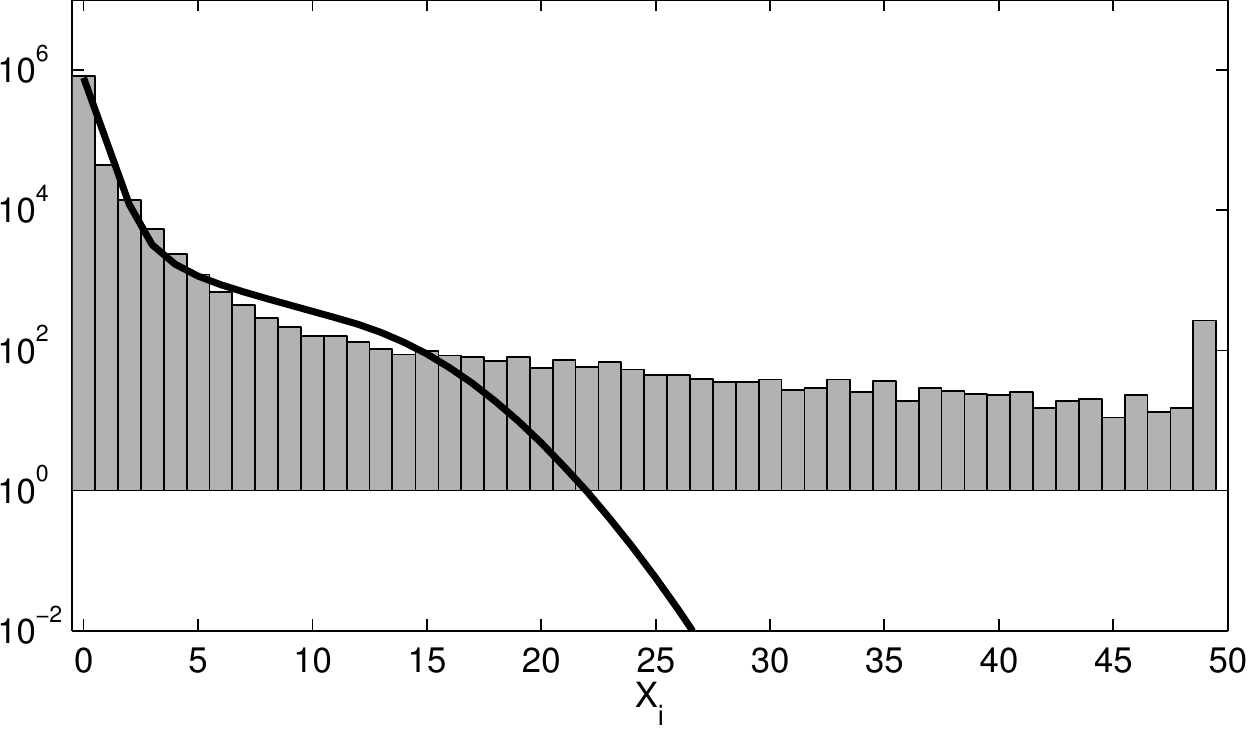} \label{fig:Hist_ML_t100}}\\
  \subfloat[\Enron, timed mix, $\tau=12h$.]{\includegraphics[width=2.3in]{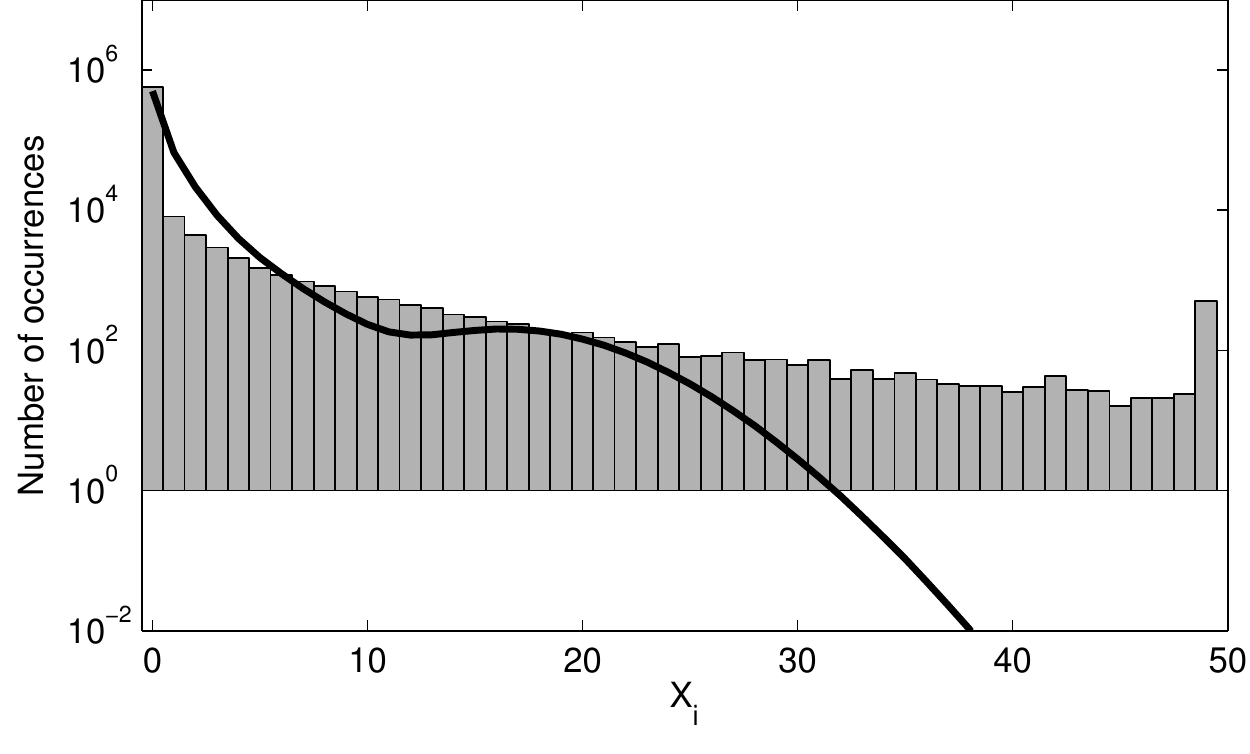} \label{fig:Hist_EN_h12}}
  \hfil
  \subfloat[\Gowalla, timed mix, $\tau=1h$.]{\includegraphics[width=2.3in]{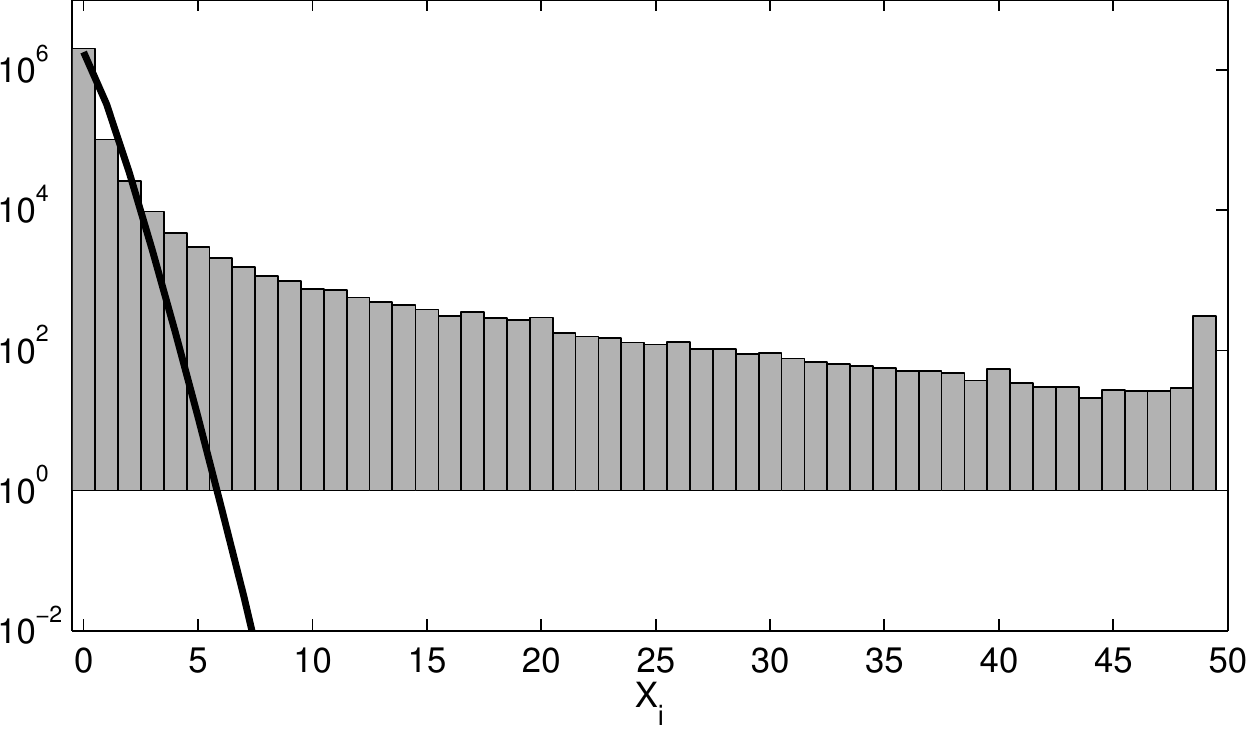} \label{fig:Hist_GO_h1}}
  \hfil
  \subfloat[\MailingList, timed mix, $\tau=24$.]{\includegraphics[width=2.3in]{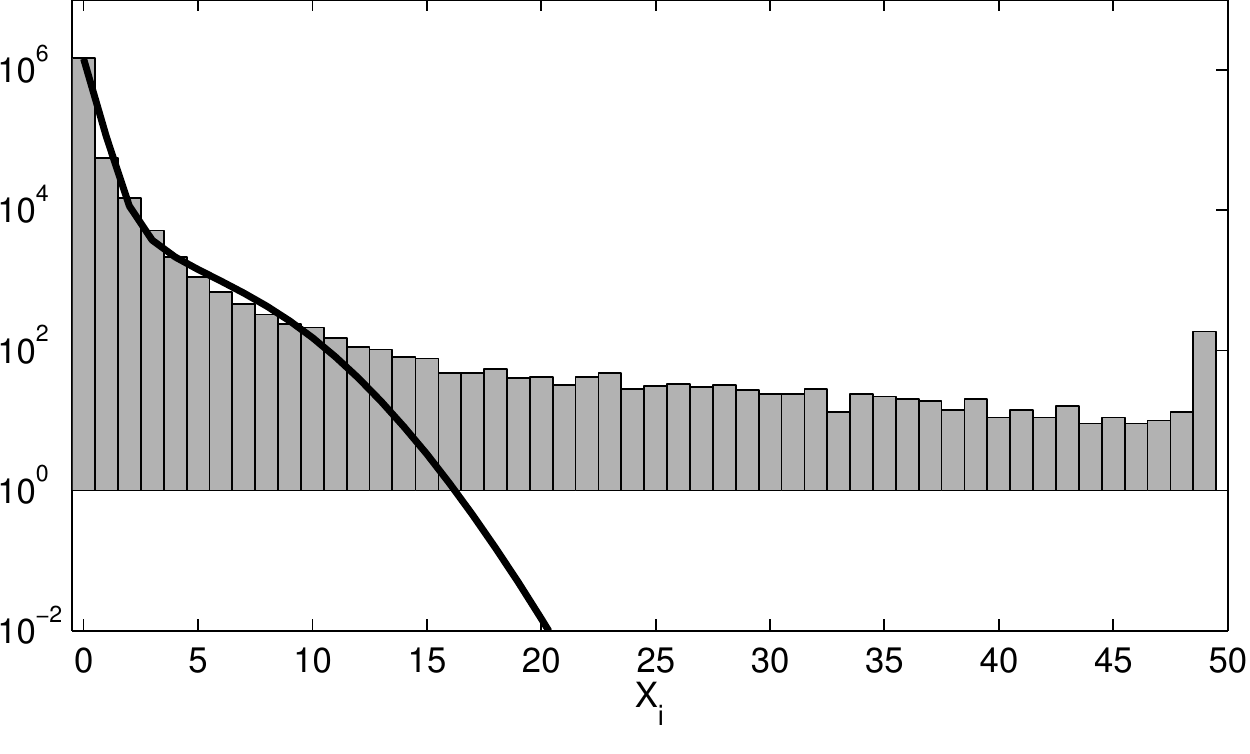} \label{fig:Hist_ML_h24}}\\
  \caption{Histograms representing the amount of messages each user sends in each round, compared to the approximation given by the theoretical models assumed in previous works (line). The last bin contains all occurrences of $X_i^r\geq 50$.}
  \label{fig:hists_input}
\end{figure*}

In Fig.~\ref{fig:hists_input}, we compare the histogram of the inputs $\{X_i^r\}$, obtained using the observations generated with our datasets, with the theoretical values given by the multinomial and Poisson models (in the threshold and the timed mixes, respectively). Here, the last bin of the histogram contains all occurrences of $X_i^r\geq50$. We conclude that the theoretical models fit the histogram for low number of messages $X_i$, but fail at capturing the large values.% the input process may take.

% \begin{figure*}[!t]
%   \centering
%   \subfloat[\Enron, threshold mix, $t=100$.]{\includegraphics[width=2.3in]{img/Hist_EN_t100} \label{fig:Hist_EN_t100}}
%   \hfil
%   \subfloat[\Gowalla, threshold mix, $t=100$.]{\includegraphics[width=2.3in]{img/Hist_GO_t100} \label{fig:Hist_GO_t100}}
%   \hfil
%   \subfloat[\MailingList, threshold mix, $t=100$.]{\includegraphics[width=2.3in]{img/Hist_ML_t100} \label{fig:Hist_ML_t100}}\\
%   \subfloat[\Enron, timed mix, $\tau=12h$.]{\includegraphics[width=2.3in]{img/Hist_EN_h12} \label{fig:Hist_EN_h12}}
%   \hfil
%   \subfloat[\Gowalla, timed mix, $\tau=1h$.]{\includegraphics[width=2.3in]{img/Hist_GO_h1} \label{fig:Hist_GO_h1}}
%   \hfil
%   \subfloat[\MailingList, timed mix, $\tau=24$.]{\includegraphics[width=2.3in]{img/Hist_ML_h24} \label{fig:Hist_ML_h24}}\\
%   \caption{Histograms representing the amount of messages each user sends in each round, compared to the approximation given by the theoretical models assumed in previous works (line). The last bin contains all occurrences of $X_i^r\geq 50$}
%   \label{fig:hists_input}
% \end{figure*}

In the analysis in this document, we do not assume a specific distribution for $\{X_i^r\}$, but consider that it is a generic \emph{stationary} process that satisfies the relation
\begin{align} 
  \Cov{X_k,X_m}&\ll \Var{X_k}\qquad 			&\forall k,m\quad k\neq m \label{eq:cond1}
  \end{align}
  and, additionally,
  \begin{align}
  \Cov{X_k,X_mX_n}&\ll \Cov{X_k^2,X_k}\qquad 		 \label{eq:cond2}\\
  \Cov{X_k^2,X_mX_n}&\ll \Cov{X_k^2,X_k^2}\qquad 	 \label{eq:cond3}
\end{align}
for all $k$, $m$, $n$ except when $k=m=n$. These assumptions mean, in other words, that the participation of a user in a given round is uncorrelated with the participation of each other user in that round. We have validated these hypotheses by computing the different sample covariances from our datasets, as shown in Table~\ref{tab:input_covs}.

 \begin{table*}[t!]
 \begin{center}
   \caption{Average values for different sample covariances of the input process in the datasets.}
  \label{tab:input_covs}
  \begin{tabular}{ r | r r r r r r }
      & \multicolumn{2}{c}{\Enron} & \multicolumn{2}{c}{\Gowalla} & \multicolumn{2}{c}{\MailingList}\\
			& $t=100$ 	& $\tau=12$ 	& $t=100$ 	& $\tau=1$ 	& $t=100$ 	& $\tau=24$ \\  \hline             
    $|\Cov{X_k,X_k}|$ 		& 7.1	& 20.1		&2.0	&2.4	&2.9 	&4.6			\\
    $|\Cov{X_k,X_m}|$ 		& 0.1	& 0.2		&0.0	&0.0	&0.0 	&0.0	\\  \rule{0pt}{3ex} 
    $|\Cov{X_k^2,X_k}|$  	& 415.9	& 5815.0	&57.2	&159.9	&171.4	&2076.1	\\
    $|\Cov{X_kX_m,X_k}|$ 	& 1.0	& 15.1		&0.3	&0.5	&0.2	&0.5	\\
    $|\Cov{X_k^2,X_m}|$ 	& 1.9	& 12.8		&0.3	&0.5	&0.5	&0.7	\\
    $|\Cov{X_kX_m,X_n}|$ 	& 0.0	& 0.3		&0.0	&0.0	&0.0	&0.0	\\ \rule{0pt}{3ex}   
    $|\Cov{X_k^2,X_k^2}|$  	& 32943.2& 2974803.8	&2412.7	&39132.0&13666.8&1381640.2	\\
    $|\Cov{X_k^2,X_kX_m}|$ 	& 30.1	& 4038.3	&6.6	&27.2	&7.0	&189.6	\\
    $|\Cov{X_k^2,X_m^2}|$  	& 51.2	& 786.2		&4.5	&10.4	&8.9	&24.8	\\
    $|\Cov{X_k^2,X_mX_n}|$ 	& 0.7	& 16.2		&0.1	&0.2	&0.1	&0.2	\\  
  \end{tabular}
  \end{center}
\end{table*}

\subsection{Modeling the output process}
\label{sec:realworld.output}
A crucial point when carrying out a performance analysis of disclosure attacks on mixes is selecting a model for the distribution $\{Y_j^r|X_1^r,\cdots,X_N^r\}$, which represents how users choose the recipients of their messages in each round. A known property of this distribution, given by the definition of sending profiles, is that $\Exp{\Y|\U}=\U\cdot\P$. However, this is true for many distributions. Every previous analysis of LSDA assumes that the choice of recipients is \emph{stationary} and that $\{Y_j^r|X_1^r,\cdots,X_N^r\}$ follows a \emph{multinomial model}, i.e.,
\begin{equation} \label{eq:dist1}
 \{Y_1^r,\cdots,Y_M^r|\U\} \sim \sum_{k=1}^N\text{Multi}\left( x_k^r, \sendprof{k}\right)\,.
\end{equation}
This model is adequate in scenarios where users choose the recipients of each of their messages in each round independently. However, when users tend to focus on a single receiver in each round, \eqref{eq:dist1} is not suitable to model the output distribution. 

In this work, we assume \emph{two models} for $\{Y_j^r|X_1^r,\cdots,X_N^r\}$ that are examples of how users can distribute their messages among the receivers while satisfying $\Exp{\Y|\U}=\U\cdot\P$:
\begin{enumerate}
 \item A \emph{multinomial model}, given by \eqref{eq:dist1}, as an example of users that cause low variance output.
 \item A \emph{maximum variance model}, given by
\begin{equation} \label{eq:dist2}
 \{Y_1^r,\cdots,Y_M^r|\U\} \sim \sum_{k=1}^N x_k^r \cdot \text{Multi}\left( 1, \sendprof{k}\right)\,.
\end{equation}
\end{enumerate}
When using these distributions, we are implicitly assuming that the choices of recipients of different senders within the same round are uncorrelated, and that the choice of recipients of the same user between rounds can be also considered uncorrelated. Our experiments in Sect.~\ref{sec:analysis.eval} confirm that the results we obtain with these approximations are accurate.

To illustrate how users' behavior changes between scenarios, we have computed the average number of recipients each sender chooses in each round of the observations generated with our datasets, as a function of the number of messages sent. This is displayed in Table~\ref{tab:recv_output}. As a reference, the average number of senders' contacts in each dataset is $125.7$ in \Enron, $16$ in \Gowalla and $9.6$ in \MailingList. These results show that users in the \Enron dataset tend to spread their messages among their contacts, behaving close to \eqref{eq:dist1}, while users in \Gowalla and \MailingList focus on a single recipient in each round, as in \eqref{eq:dist2}. 

 \begin{table}[t!]
 \begin{center}
   \caption{Average number of recipients chosen by the senders in each round, as a function of the number of messages sent.}
  \label{tab:recv_output}
  \begin{tabular}{ c r | r r r r r }
   \multicolumn{2}{c|}{\#~messages ($X_i$)} & $=2$	& $=3$	& $=4$ 	& $=5$ & $\geq6$ 	\\ \hline
   \multirow{2}{*}{\Enron} & $t=100$ 	& 1.85 &   2.71  &  3.53  &  4.40 &  13.56	\\ 
    & $\tau=12h$ & 1.85 &   2.69  &  3.49  &  4.32 &  14.02	\\ \rule{0pt}{2.8ex} 
    \multirow{2}{*}{\Gowalla} & $t=100$  & 1.03 &   1.05  &  1.06  &  1.08 &   1.11	\\
   & $\tau=1h$& 1.10 &   1.14  &  1.17  &  1.18 &   1.26	\\ \rule{0pt}{2.8ex} 
    \multirow{2}{*}{\MailingList}& $t=100$ & 1.29  &  1.46 &   1.53 &   1.53 &   1.57	\\
    & $\tau=24h$ & 1.28 &   1.49  &  1.56  &  1.55   & 1.71 \\
  \end{tabular}
  \end{center}
\end{table}

%% file: sections/analysis.tex
\section{Extended Performance Analysis of the Least Squares Disclosure Attack}
\label{sec:analysis}

We now assess the profiling accuracy of the Least Squares Disclosure Attack with the assumptions in the input and output processes proposed in the previous section, which we have validated with traffic from real-world scenarios. The profiling accuracy is measured as the Mean Squared Error (MSE) between the attacker's estimation of the \emph{sending profiles} of the users and their real values, i.e., $\MSEi\doteq\sum_{j=1}^N |\prob{j}{i}-\pest{j}{i}|^2$. This analysis generalizes previous ones~\cite{PETS12,WIFS12,GlobalSIP,PETS14}, accommodating different types of mixes and being able to model real-world behavior, at the expense of accuracy.

\subsubsection{Theoretical approximation of the average MSE}
\label{sec:analysis.MSE}

Our goal is to obtain an approximation of the \emph{average} $\MSEi$ when using \eqref{eq:LSDA} to estimate the sending profiles, where this average is computed over all the realizations of $\U$ and $\Y$ obtained with users' average behavior $\P$. For simplicity, we 	omit the conditioning on $\P$ in the derivations below.

For the analysis in this section, we introduce additional notation regarding the statistics of the input and output processes. We use $\mu(i)$ to refer to the expected value of $X_i$, and $\mu_n(i)$ is its $n$th central moment. Vector $\muv$ contains all $\mu(i)$ for each sender, i.e., $\muv\doteq[\mu(1),\cdots,\mu(N)]^T$. Matrix $\M$ contains these values arranged in its main diagonal, i.e., $\M\doteq\text{diag}\{\mu(1),\cdots,\mu(N)\}$ and, similarly, $\M_n\doteq\text{diag}\{\mu_n(1),\cdots,\mu_n(N)\}$. We use the parameter $s_{j,i}\doteq\prob{j}{i}(1-\prob{j}{i})$, which is closely related to the variance of the outputs, and the diagonal matrix $\Sj\doteq\text{diag}\{s_{j,1},\cdots,s_{j,N}\}$. Finally, we define the \emph{uniformity} of the sending profile of user $i$ as $\uniformi{i}\doteq1-\sum_{j=1}^M \prob{j}{i}^2$. The uniformity gives an idea of how random the behavior of a user is, and ranges from 0, when sender only has one contact, to $(M-1)/M$, when this user sends messages to all the receivers with the same probability during the observation period. Note that $\sum_{j=1}^M s_{j,i}=\uniformi{i}$. 

We start the derivations by showing that the LSDA estimator is unbiased. This is straightforward from the fact that, given a matrix of input messages $\U$ and the average behavior of the senders $\P$, the expected value of the output is
\begin{equation} \label{eq:ExpY}
 \Exp{\Y|\U}=\U\cdot \P
\end{equation}
where $\Exp{\cdot}$ is taken along all the possible assignments of the messages in $\U$ to the receivers, following $\P$. Using \eqref{eq:ExpY} together with \eqref{eq:LSDA}, we get $\text{E}\{\Pest\}=\P$ (alternatively, $\Exp{\pjest{j}}=\pj{j}$). This property allows to write, using the law of total variance,
\begin{equation} \label{eq:sigmaYj1}
 \mathbf{\Sigma}_{\pjest{j}}=\Exp{\mathbf{\Sigma}_{\pjest{j}|\U}}=\Exp{(\U^T\U)^{-1}\U^T\mathbf{\Sigma}_{\Yj{j}|\U}\U(\U^T\U)^{-1}}
\end{equation}
where $\mathbf{\Sigma}_{\Yj{j}|\U}\doteq\Exp{(\Yj{j}-\Exp{\Yj{j}|\U})(\Yj{j}-\Exp{\Yj{j}|\U})^T|\U}$.

Since we have assumed that the input process is stationary, using the Law of Large Numbers and considering that the number of rounds observed $\rho$ is large enough, we approximate
\begin{equation}
 \lim\limits_{\rho\rightarrow\infty} \U^T\U/\rho\rightarrow \Rx
\end{equation}
where $\Rx$ is the autocorrelation matrix of the input process, i.e., an $N\times N$ symmetric matrix whose $m,n$th element is $\Exp{X_m X_n}$. Using \eqref{eq:cond1}, we write this matrix as
\begin{equation} \label{eq:Rx}
 \Rx\approx\muv\muv^T+\MM.
\end{equation}
The inverse of \eqref{eq:Rx} can be computed applying the Sherman-Morrison formula \cite{shermanmorrison}, which gives us
\begin{equation} \label{eq:Rxinv}
 \Rx^{-1}\approx\MM^{-1}\left(\mathbf{I}_N-\gamma \muv\muv^T \MM^{-1}\right)
\end{equation}
where $\gamma\doteq1/(1+\muv^T\MM^{-1}\muv)$.
Therefore, when the number of rounds observed is large, \eqref{eq:sigmaYj1} can be approximated as
\begin{equation} \label{eq:sigmaYj}
 \mathbf{\Sigma}_{\pjest{j}}\approx \frac{1}{\rho} \Rx^{-1} \Rxyx \Rx^{-1}.
\end{equation}
where the middle term is $\Rxyx\doteq\frac{1}{\rho}\Exp{\U^T\mathbf{\Sigma}_{\Yj{j}|\U}\U}$. In order to compute the covariance  matrix $\mathbf{\Sigma}_{\Yj{j}|\U}$, we analyze separately the two scenarios \eqref{eq:dist1} and \eqref{eq:dist2} we consider for the distribution of the output process given the inputs.

\paragraph{Multinomial model}
Using \eqref{eq:dist1} together with our assumptions, we approximate the middle term of \eqref{eq:sigmaYj} as
\begin{equation} \label{eq:mid1}
  \Rxyx\approx\left(\sum_{k=1}\mu(k)s_{j,k}\right)\cdot \left(\muv\muv^T+\MM\right)+\MMM \Sj.
\end{equation}
Finally, plugging \eqref{eq:Rxinv} and \eqref{eq:mid1} into \eqref{eq:sigmaYj} and performing matrix multiplications we obtain $\mathbf{\Sigma}_{\pjest{j}}$. Then, taking the $i$-th diagonal element of this matrix, which is $\Var{\pest{j}{i}}$, adding this element along $j$, and further considering $\sum_{k=1, k\neq i}^N \mu^2(k)/\mu_2(k) \gg 0$ for all $i\in\{1,\cdots,N\}$, we obtain:
\begin{equation} \label{eq:MSEmin}
  \MSEimin\approx \frac{1}{\rho} \cdot \frac{1}{\mu_2(i)}\left( \sum_{k=1}^N \mu(k)\cdot\uniformi{k} + \frac{\mu_3(i)}{\mu_2(i)}\cdot \uniformi{i}\right) 
\end{equation}

\paragraph{Maximum variance model}
We now analyze the performance of the LSDA estimator when the output distribution is \eqref{eq:dist2}. In that case, the middle term of \eqref{eq:sigmaYj} becomes

\begin{equation} \label{eq:mid2}
 \Rxyx\approx\left(\sum_{k=1}^N(\mu(k)^2+\mu_2(k))s_{j,k}\right)\cdot \left(\muv\muv^T+\MM\right)+\MMMM \Sj.
\end{equation}
Operating as explained before to obtain the MSE in the estimation of the sending profile of user $i$, we get
\begin{equation} \label{eq:MSEmax}
  \MSEimax\approx \frac{1}{\rho} \cdot \frac{1}{\mu_2(i)}\left( \sum_{k=1}^N\left(\mu(k)^2+\mu_2(k)\right)\cdot\uniformi{k} + \frac{\mu_4(i)}{\mu_2(i)}\cdot \uniformi{i}\right) 
\end{equation}

The formulas \eqref{eq:MSEmin} and \eqref{eq:MSEmax} provide new insights into how LSDA's error depends on the system parameters. This error decreases with $\rho$, since it becomes easier for the attacker to estimate the behavior of the users as more observations are available. The \emph{variance of the input process} $X_i$ \emph{decreases} the estimation error of $\sendprof{i}$, i.e., it is easier to separate the sending behavior of a user from the others when we have rounds where that user participates a lot as well as rounds where that user is not present. The $\MSEi$ also \emph{increases} with the contribution of all senders to the \emph{output variance}, more strongly when users behave as in \eqref{eq:dist2} than as in \eqref{eq:dist1}. The role of the uniformity of the profiles $\uniformi{k}$ in the MSE is also very relevant: estimating the sending profiles is a much easier task when users only contact very few receivers (i.e., low $\uniformi{k}$) than when they distribute their messages among a larger population (i.e., $\uniformi{k}$ close to 1).

\subsubsection{Evaluation}
\label{sec:analysis.eval}

We now evaluate our formulas, applying LSDA to the anonymized traces of real traffic. For each dataset, mix configuration, and number of rounds observed $\rho\in\{0.1\rho_{max}, 0.2\rho_{max}, \cdots, \rho_{max}\}$, where $\rho_{max}$ is the total number of rounds in the observations, we perform LSDA and compute the real $\MSEi$. We then represent the average $\MSEi$ of those users $i$ that meet three conditions: they are among the $40\%$ most active users, they belong to the $40\%$ users that remain active for the largest number of rounds, and furthermore they participate before $0.3\rho_{max}$ rounds have been observed. We do this to avoid sporadic peaks in the average $\MSEi$, which are the result of estimating the sending profile of a user that barely participates in the system, and to be able to see the trend of the MSE with clarity. 

\begin{figure*}[!t]
  \centering
  \subfloat[\Enron, threshold mix, $t=100$.]{\includegraphics[width=2.3in]{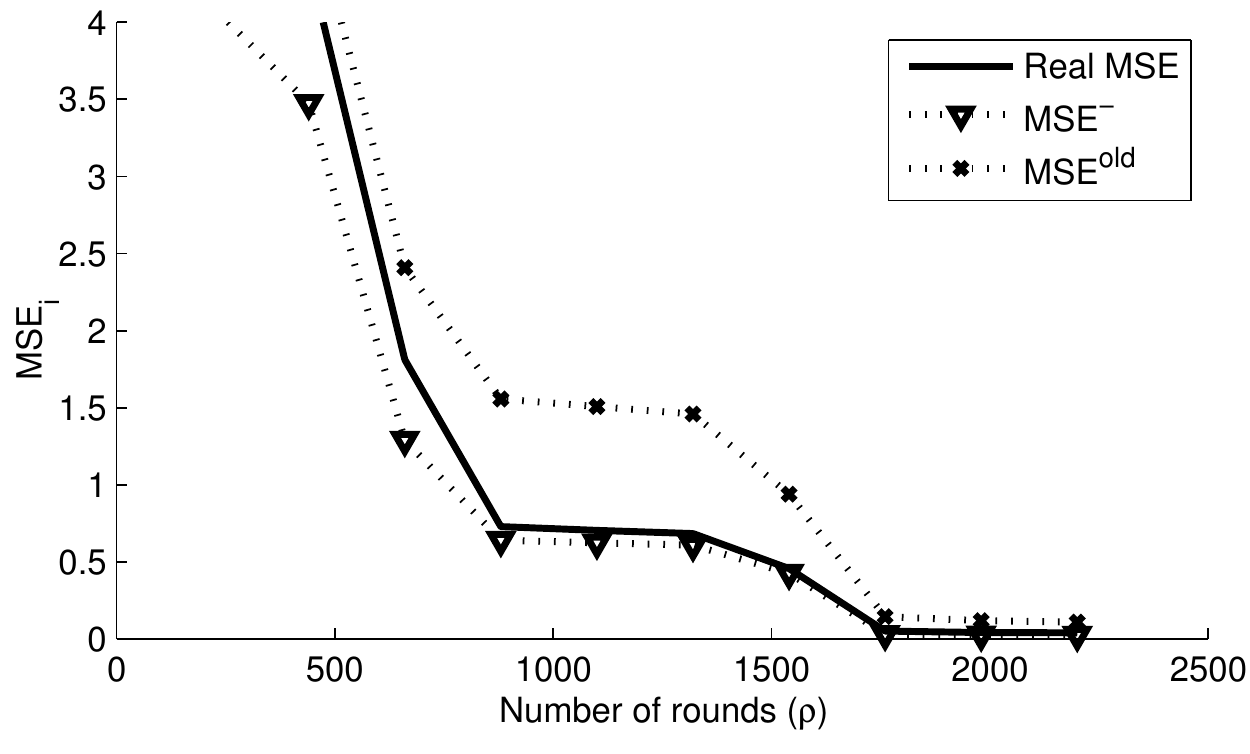} \label{fig:MSE_EN_t_4}}
  \hfil
  \subfloat[\Gowalla, threshold mix, $t=100$.]{\includegraphics[width=2.3in]{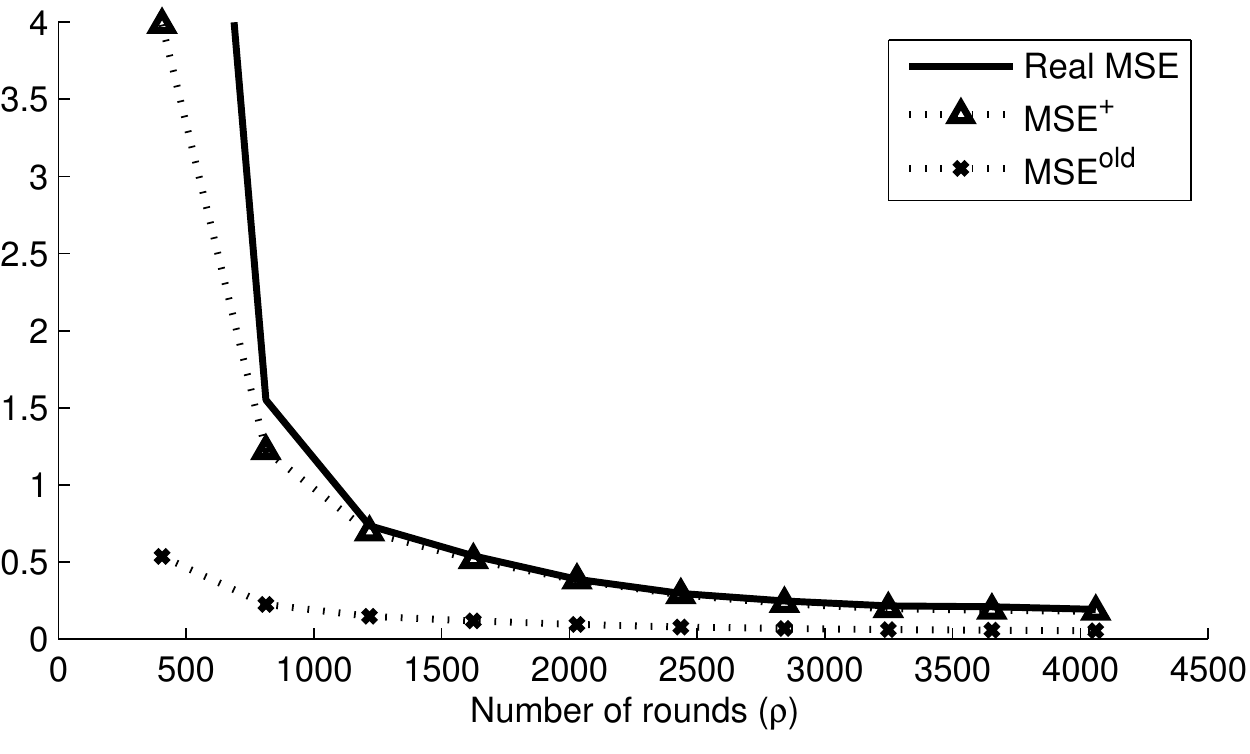} \label{fig:MSE_GO_t_4}}
  \hfil
  \subfloat[\MailingList, threshold mix, $t=100$.]{\includegraphics[width=2.3in]{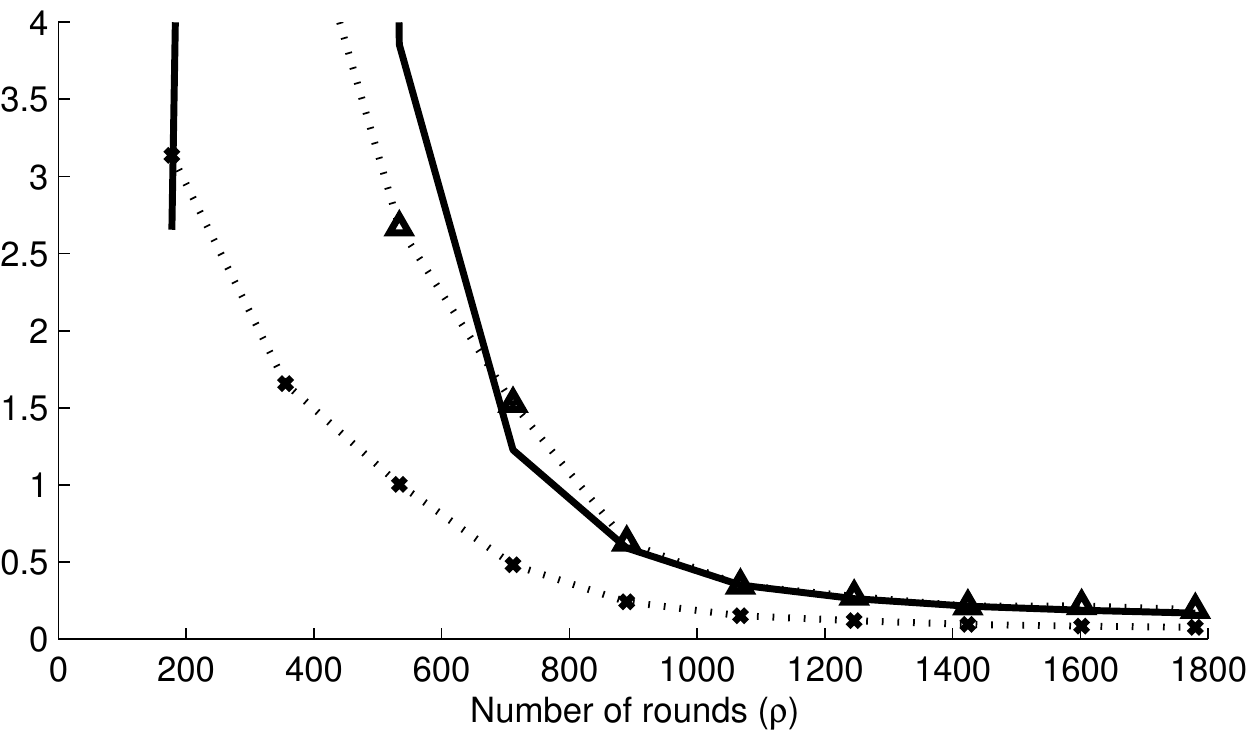} \label{fig:MSE_ML_t_4}}\\
  
  \subfloat[\Enron, timed mix, $\tau=12h$.]{\includegraphics[width=2.3in]{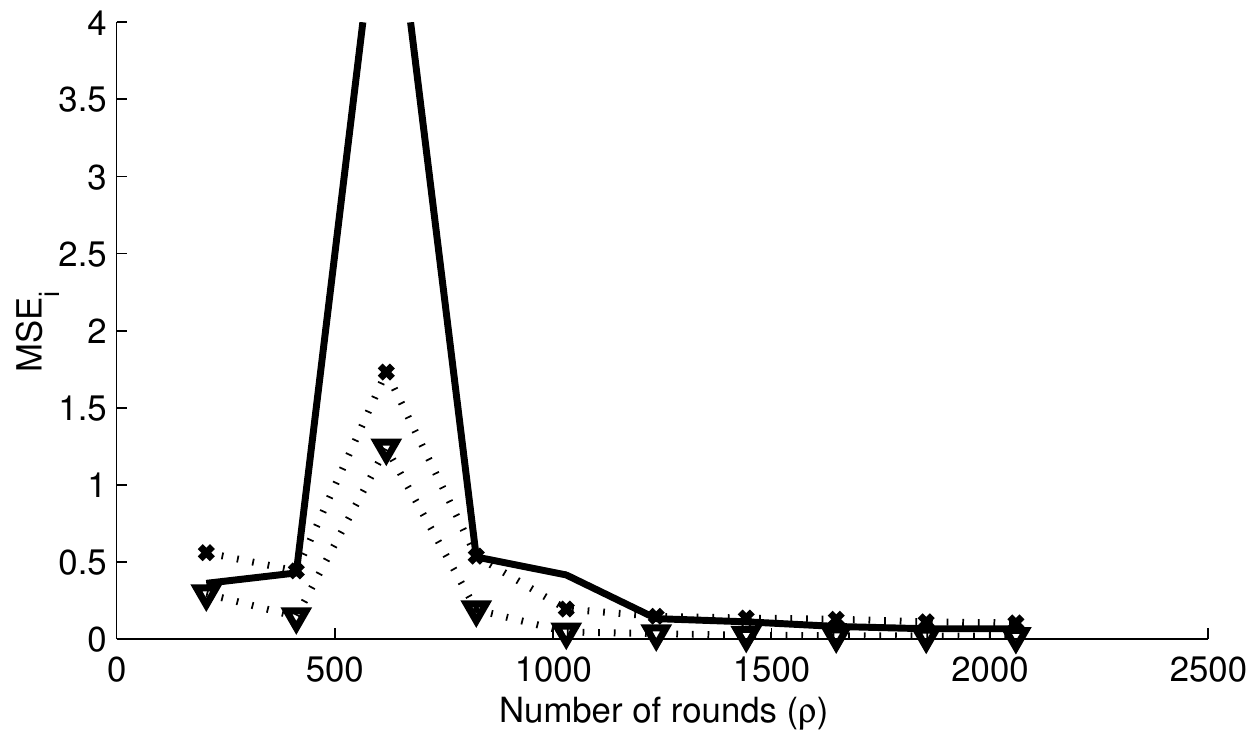} \label{fig:MSE_EN_h_4}}
  \hfil
  \subfloat[\Gowalla, timed mix, $\tau=1h$.]{\includegraphics[width=2.3in]{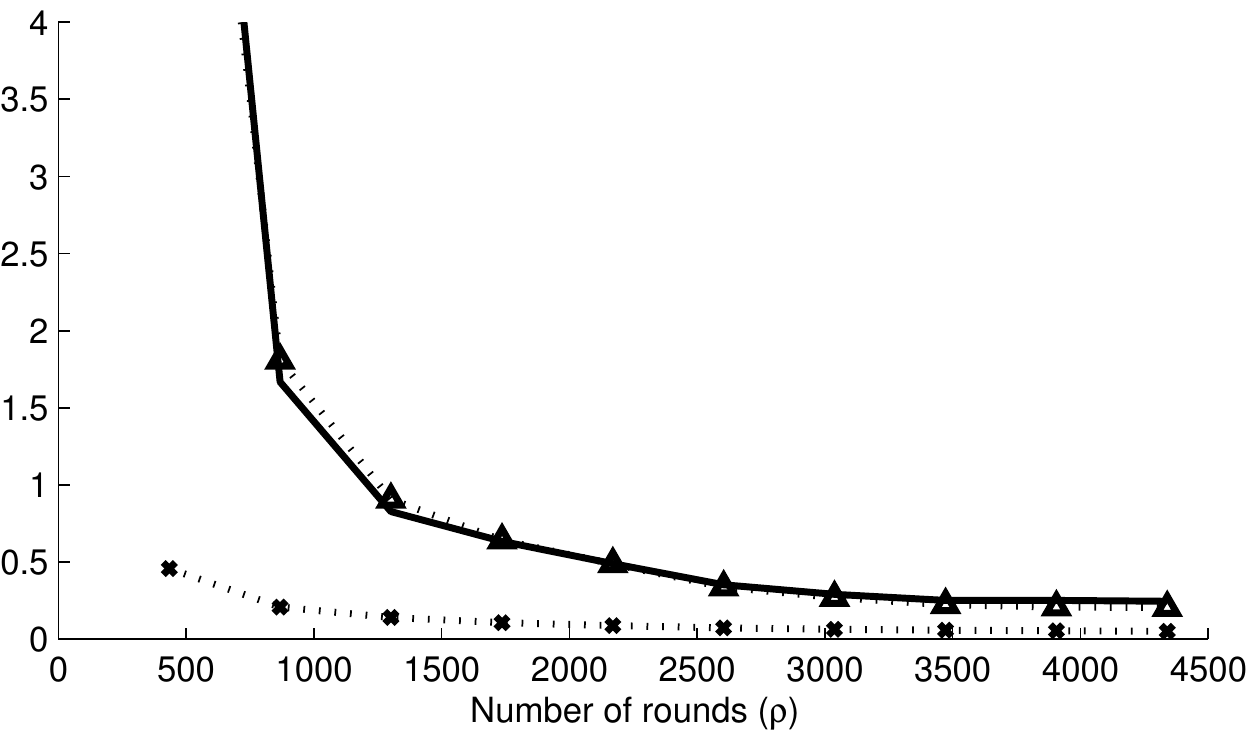} \label{fig:MSE_GO_h_4}}
  \hfil
  \subfloat[\MailingList, timed mix, $\tau=24h$.]{\includegraphics[width=2.3in]{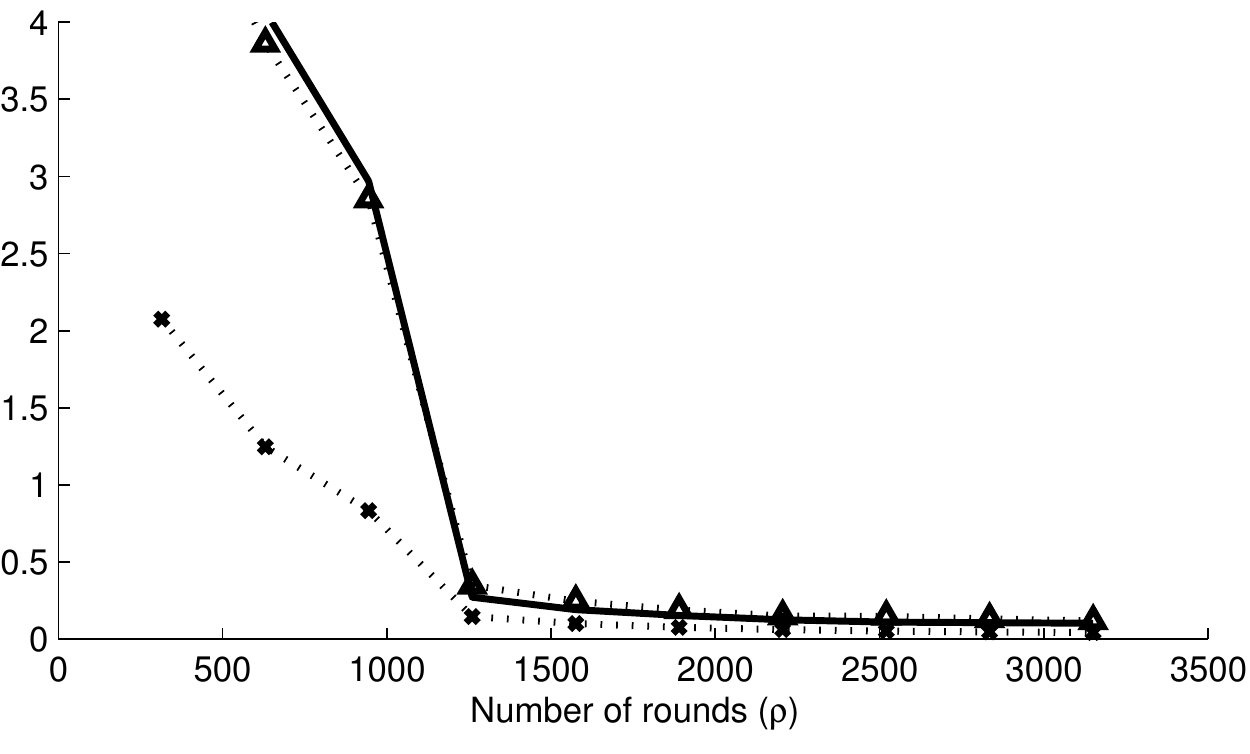} \label{fig:MSE_ML_h_4}}\\
  
  \caption{Average $\MSEi$ evolution with $\rho$ using the \Enron, \Gowalla and \MailingList datasets, for different types of mixes.}
  \label{fig:all_results}
\end{figure*}

Figure~\ref{fig:all_results} shows this average $\MSEi$ for the \Enron, \Gowalla and \MailingList datasets, together with the theoretical formulas $\MSEimin$ and $\MSEimax$ in \eqref{eq:MSEmin} and \eqref{eq:MSEmax}. We only plot the theoretical approximation that better suits each scenario: $\MSEimin$ in the \Enron experiments and $\MSEimax$ in the \Gowalla and \MailingList experiments. We also plot the theoretical MSE from previous works, denoted by $\text{MSE}^{\texttt{old}}$, which has been taken from \cite{GlobalSIP} and \cite{PETS14} for the threshold and the timed mix experiments, respectively. We set the limits of the vertical axis to the same value in all figures to ease the comparison between them. These limits make early values of the MSE (low $\rho$) to fall outside the plot, but allow to see with more detail the performance of the attack for large values of $\rho$. We do this on purpose: we are predicting the asymptotic $\MSEi$ of the attack, so the results for low values of $\rho$ are not significant in our evaluation.

We see that our approximations improve those given by previous work, especially in those scenarios where the multinomial model for the choice of recipients is not appropriate (\Gowalla and \MailingList). We note that the number of rounds we can generate with the \Enron database in Fig.~\ref{fig:MSE_EN_h_4} is not large enough to appreciate this improvement, due to the spike we observe in that experiment at early values of $\rho$. This sudden increase of the MSE, as well as the one in Fig.~\ref{fig:MSE_ML_t_4}, happens for two reasons: first, when the number of rounds observed is small, it is easier for the matrix $\U^T\U$ to be ill-conditioned, which results in a poor estimation of the sending profiles (cf.~\eqref{eq:LSDA}), and therefore in a large MSE in the realization. This spike is not predicted by our theoretical formulas, since they approximate the \emph{average} MSE. On the other hand, in the \Enron dataset, most of the users whose $\MSEi$ we average start sending messages when the adversary has observed around $30\%$ of the total number of rounds. This causes an increase in the $\MSEi$ at around $\rho=600$, as we are adding users to the average $\MSEi$ that have barely participated in the system. The average $\MSEi$ stabilizes as the number of rounds observed increases since the number of users used for the computation of the average $\MSEi$ we represent remains unchanged.

In all cases, the $\MSEi$ decreases as the number of observed rounds $\rho$ increases, as predicted by our formulas, except for the spikes in Figs.~\ref{fig:MSE_EN_h_4} and \ref{fig:MSE_ML_t_4} whose origin we have already explained. Due to these spikes, comparing the results of the experiments in the \Enron and \MailingList datasets is not possible. However, we can see that the MSE in the experiments with \Gowalla is stable, and always larger in the threshold mix scenario (Fig.~\ref{fig:MSE_GO_t_4}). The reason for this is the following: the variance of the the input process in a threshold mix is smaller than that in a timed mix for the same average number of messages sent per round. This is the the case in the \Gowalla experiments, since the number of rounds we generate in the threshold and timed mix experiments is approximately the same. As predicted by our theoretical formulas, a system with lower input variance provides more protection against the LSDA attacker.

%% file: sections/conclusions.tex
\section{Conclusions}
\label{sec:conclusions}

We have analyzed the effects of real-world user behavior in the performance of the least squares disclosure attack~\cite{PETS12} in mix-based anonymous communication systems, considering mixes that do not delay messages between communication rounds. To validate our work, we have obtained real traffic observations from three publicly available datasets of different nature: emails sent between the employees of a company, location check-ins from an online social network, and users' posts to mailing lists. By studying these data, we confirm that the hypotheses upon which former analyses of the least squares disclosure attack are based~\cite{PETS12,WIFS12,PETS14} are not adequate to model real-world behavior, and hence we formulate new ones. Based on these new assumptions, we develop a generalized performance analysis of the attack, which we validate with our datasets, confirming that it accurately models the estimation error of the attacker in the considered realistic scenarios. This analysis accommodates a wide variety of mix and users' behavior, and provides new insights into the statistics that affect the protection of the users: the variability in the participation of the users in the system contributes to the attacker's success, while the variability in the messages received by users worsens the attacker's estimation.

%This work is a step forwards towards the analysis of disclosure attacks in realistic scenarios. As a subject for future work, it would be interesting to consider more complex mixing strategies, such as mixes that delay messages between rounds.

%% file: main.bbl
\begin{thebibliography}{10}
\providecommand{\url}[1]{#1}
\csname url@samestyle\endcsname
\providecommand{\newblock}{\relax}
\providecommand{\bibinfo}[2]{#2}
\providecommand{\BIBentrySTDinterwordspacing}{\spaceskip=0pt\relax}
\providecommand{\BIBentryALTinterwordstretchfactor}{4}
\providecommand{\BIBentryALTinterwordspacing}{\spaceskip=\fontdimen2\font plus
\BIBentryALTinterwordstretchfactor\fontdimen3\font minus
  \fontdimen4\font\relax}
\providecommand{\BIBforeignlanguage}[2]{{%
\expandafter\ifx\csname l@#1\endcsname\relax
\typeout{** WARNING: IEEEtran.bst: No hyphenation pattern has been}%
\typeout{** loaded for the language `#1'. Using the pattern for}%
\typeout{** the default language instead.}%
\else
\language=\csname l@#1\endcsname
\fi
#2}}
\providecommand{\BIBdecl}{\relax}
\BIBdecl

\bibitem{SDA}
G.~Danezis, ``Statistical disclosure attacks: Traffic confirmation in open
  environments,'' in \emph{Proceedings of Security and Privacy in the Age of
  Uncertainty}, Gritzalis, Vimercati, Samarati, and Katsikas, Eds., {IFIP
  TC11}.\hskip 1em plus 0.5em minus 0.4em\relax Athens: Kluwer, May 2003, pp.
  421--426.

\bibitem{MD04}
N.~Mathewson and R.~Dingledine, ``Practical traffic analysis: Extending and
  resisting statistical disclosure,'' in \emph{4th Workshop on Privacy
  Enhancing Technologies}, ser. LNCS, D.~Martin and A.~Serjantov, Eds., vol.
  3424.\hskip 1em plus 0.5em minus 0.4em\relax Springer, 2004, pp. 17--34.

\bibitem{PMDA}
C.~Troncoso, B.~Gierlichs, B.~Preneel, and I.~Verbauwhede, ``Perfect matching
  disclosure attacks,'' in \emph{8th Symposium on Privacy Enhancing
  Technologies}, ser. LNCS, N.~Borisov and I.~Goldberg, Eds., vol. 5134.\hskip
  1em plus 0.5em minus 0.4em\relax Springer-Verlag, 2008, pp. 2--23.

\bibitem{Vida}
G.~Danezis and C.~Troncoso, ``Vida: How to use {Bayesian} inference to
  de-anonymize persistent communications,'' in \emph{9th Privacy Enhancing
  Technologies Symposium}, ser. LNCS, I.~Goldberg and M.~J. Atallah, Eds., vol.
  5672.\hskip 1em plus 0.5em minus 0.4em\relax Springer, 2009, pp. 56--72.

\bibitem{PETS12}
F.~P\'erez-Gonz\'alez and C.~Troncoso, ``Understanding statistical disclosure:
  A least squares approach,'' in \emph{Privacy Enhancing Technologies - 12th
  Symposium}, ser. LNCS, vol. 7384.\hskip 1em plus 0.5em minus 0.4em\relax
  Springer-Verlag, 2012, pp. 38--57.

\bibitem{WIFS12}
F.~P{\'e}rez-Gonz{\'a}lez and C.~Troncoso, ``A least squares approach to user
  profiling in pool mix-based anonymous communication systems,'' in \emph{IEEE
  Workshop on Information Forensics and Security}, 2012, pp. 115--120.

\bibitem{PETS14}
S.~Oya, C.~Troncoso, and F.~P{\'e}rez-Gonz{\'a}lez, ``Do dummies pay off?
  limits of dummy traffic protection in anonymous communications,'' in
  \emph{14th Symposium on Privacy Enhancing Technologies}, 2014.

\bibitem{GlobalSIP}
S.~Oya, C.~Troncoso, and F.~P\'erez-Gonz\'alez, ``Meet the family of
  statistical disclosure attacks,'' \emph{IEEE Global Conference on Signal and
  Information Processing}, p.~4p, 2013.

\bibitem{WPES13}
G.~Danezis and C.~Troncoso, ``You cannot hide for long: De-anonymization of
  real-world dynamic behaviour,'' in \emph{Proceedings of the 12th ACM Workshop
  on Workshop on Privacy in the Electronic Society}, ser. WPES '13.\hskip 1em
  plus 0.5em minus 0.4em\relax ACM, 2013, pp. 49--60.

\bibitem{malinka2009analyses}
K.~Malinka, P.~Han{\'a}{\v{c}}ek, and D.~Cvr{\v{c}}ek, ``Analyses of real email
  traffic properties,'' \emph{Radioengineering}, vol.~18, no.~4, p.~7, 2009.

\bibitem{shermanmorrison}
J.~Sherman and W.~J. Morrison, ``Adjustment of an inverse matrix corresponding
  to a change in one element of a given matrix,'' \emph{The Annals of
  Mathematical Statistics}, vol.~21, no.~1, pp. 124--127, 1950.

\end{thebibliography}
